\documentclass[10pt,conference,compsocconf]{IEEEtran}
\usepackage{times}
\usepackage{url}
\usepackage{tikz}

\usepackage{caption}
\usepackage{threeparttable}
\usepackage{tabularx}
\usepackage{multirow}
\usepackage{hhline}

\ifCLASSINFOpdf
\else
\fi
\usepackage[cmex10]{amsmath}

\usepackage{listings}
\usepackage{color}

\definecolor{mygreen}{rgb}{0,0.5,0}
\definecolor{mygray}{rgb}{0.9,0.9,0.9}
\definecolor{mymauve}{rgb}{0.58,0,0.82}
\definecolor{myblue}{rgb}{0.0,0.0,0.5}

\begin{document}
\pagenumbering{gobble}
%
\title{\textbf{\Large On the Resilience of a QKD Key  
\\[-1.5ex]Synchronization Protocol for IPsec }\\[0.2ex]}

\author{
\IEEEauthorblockN{~\\[-0.4ex]\large Stefan Marksteiner\\[0.3ex]\normalsize}
\IEEEauthorblockA{
	JOANNEUM RESEARCH GmbH\\
	DIGITAL - Institute for Information \\
	and Communication Technologies\\
	Graz, Austria\\ 
	Email: stefan.marksteiner@joanneum.at
}
\and
\IEEEauthorblockN{~\\[-0.4ex]\large Benjamin Rainer\\[0.3ex]\normalsize}
\IEEEauthorblockA{
	University of Klagenfurt\\
	Institute of Information Technology\\
	Multimedia Communication Group\\
	Klagenfurt, Austria\\
	Email: benjamin.rainer@itec.aau.at}
\and
\IEEEauthorblockN{~\\[-0.4ex]\large Oliver Maurhart\\[0.3ex]\normalsize}
\IEEEauthorblockA{
	AIT Austrian Institute of Technology GmbH\\
	Digtal Safety \& Security Department\\
	Optical Quantum Technology\\
	Klagenfurt, Austria\\
	Email: oliver.maurhart@ait.ac.at}
}


\maketitle

\begin{abstract}
This paper presents a practical solution to the problem of limited bandwidth in Quantum Key Distribution (QKD)-secured
communication through using rapidly rekeyed Internet Protocol security (IPsec) links. QKD is a cutting-edge security
technology that provides mathematically proven security by using quantum physical effects and information theoretical
axioms to generate a guaranteed non-disclosed stream of encryption keys. Although it has been a field of theoretical
research for some time, it has only been producing market-ready solutions for a short period of time. The downside of
this technology is that its key generation rate is only around 52,000 key bits per second over a distance of 50 km.  As
this rate limits the data throughput to the same rate, it is substandard for normal modern communications, especially
for securely interconnecting networks. IPsec, on the other hand, is a well-known security protocol that uses classical
encryption and is capable of exactly creating site-to-site virtual private networks. This paper presents a solution
that combines the performance advantages of IPsec with QKD. The combination sacrifices only a small portion of QKD
security by using the generated keys a limited number of times instead of just once. As a part of this, the solution
answers the question of how many data bits per key bit make sensible upper and lower boundaries to yield high
performance while maintaining high security. While previous approaches complement the Internet Key Exchange protocol
(IKE), this approach simplifies the implementation with a new key synchronization concept, proposing a lightweight
protocol that uses relatively few, slim control messages and sparse acknowledgement. Furthermore, it provides a
Linux-based module for the AIT QKD software using the Netlink XFRM Application Programmers Interface to feed the
quantum key to the IPsec cipher. This enables wire-speed, QKD-secured communication links for business applications.
This paper, apart from the description of the solution itself, describes the surrounding software environment,
including the key exchange, and illustrates the results of thorough test simulations with a variety of different protocol parameter settings.
\end{abstract}


\begin{IEEEkeywords}
Quantum Key Distribution; QKD; IPsec; Cryptography; Security; Networks.%
\end{IEEEkeywords}

%
\IEEEpeerreviewmaketitle

\section{Introduction and Motivation}

A recent paper presents an
approach to combine quantum key distribution (QKD) with IPsec by using QKD to provide IPsec with the
cryptographic keys necessary for its operation \cite{MM:2015}. This article extends the work described in the mentioned
paper such that it further examines the impact of noise (and other effects that are likely to happen in real-world
networks) on the presented solution.
Quantum cryptography, in this particular case quantum key distribution, has the purpose to ensure the
confidentiality of a communication channel between two parties. The major difference to classical cryptography is that it
does not rely on assumptions about the security of the mathematical problem it is based on, nor the computing power of
a hypothetical attacker. Instead, QKD presents a secure method of exchanging keys by connecting the two communicating
parties with a quantum channel and thereby supplying them with guaranteed secret and true random key material
\cite[p.743]{ZBGR:1998}. When the key is applied through a Vernam cipher (also called one time pad - OTP) on a data
channel on any public network, this method provides the channel with information-theoretically (in other words mathematically
proven) security \cite[p.583]{NC:2000}. An \textit{information-theoretically secure}\footnote{Shannon used the term
\protect\textit{secrecy} instead of security. In cryptography, more secrecy means more security
\protect\cite[p.1]{ZBGR:1998}. Thus, the two terms are synonymous in this context.} system means, besides a mathematical
proof, that this system is still secure if an attacker has infinite resources and time at his disposal to
cryptographically analyze it \cite[pp.659]{Shannon:1949}. The downside of combining QKD with OTP is the limitation to
approximately fifty-two kilobits over fifty kilometers, shown in a practical QKD setup \cite[p.1]{WHHLPZ:2015}, due to
physical and technical factors, since in OTP one key bit is consumed by one data bit \cite[S.9]{TPHFLQMHZ:2005}. OTP
is so far the only known information-theoretically (also called unconditionally) secure encryption algorithm \cite[pp.177 - 178]{SK:2010}.
The offered data rate, however, does not meet the requirements of modern communications. Another practical approach came
to the same conclusion and therefore uses the Advanced Encryption Standard (AES) instead of OTP \cite[p.6]{XCWYZLZZLLHG:2009}.
As IPsec is a widespread security protocol suite that provides integrity, authenticity and confidentiality for data
connections, this approach uses the combination of IPsec and QKD to overcome this restrictions \cite[p.4]{RFC4301}.

To save valuable key material, this solution uses it for more than one data packet in
IPsec, thus increasing the effective data rate, which is thereby not limited to the key rate anymore. 
Furthermore, using this approach, the presented solution benefits from the flexibility of IPsec in terms of
cryptographic algorithms and cipher modes. In contrast to
most of the previous approaches (see Section \ref{sec:Previous}), that supplemented the Internet Key Exchange (IKE)
protocol or combine in some way quantum-derived and classical keys, this paper refrains from using IKE (for a key exchange is rather the objective of
QKD, as described later) in favor of a specialized, lightweight key synchronization protocol, working with a
master/slave architecture. The goal of this protocol is to achieve very high changing rates of purely quantum-derived keys on the
communicating peers while maintaining the keys synchronous in a very resilient manner, which means to deal with
suboptimal networking conditions including packet losses and late or supplicate packets. In order to fulfill this
objective, the following questions need to be clarified:
\begin{itemize}
  \item What is the minimum acceptable frequency of changing the IPsec key that will ensure sufficient
  security?
  \item What is the maximum acceptable frequency of changing the IPsec key to save QKD key
  material?
  \item Is the native Linux kernel implementation suitable for this task?
  \item How can key synchronicity between the communication peers be assured at key periods of 50 milliseconds and less?
\end{itemize}
As a proof of concept, this paper further presents a software solution, called QKDIPsec, implementing this approach in
C++. This software is intended to be used as an IPsec module for the multi platform
hardware-independent AIT QKD software, which provides already a market-ready
solution for OTP-based QKD. The module achieves over forty key changes per second for the IPsec subsystem
within the Linux kernel. At present time, the software uses a static key ring buffer for testing purposes instead of
actual QKD keys, for the integration of QKDIPsec into the AIT QKD software is yet to be implemented (although most of
the necessary interfaces are already present). The ultimate goal is to deliver a fully operational IPsec module for the
AIT QKD software.

The following Section \ref{sec:Previous} of this paper describes previous approaches on combining IPsec and QKD.
Section \ref{sec:keychange} describes considerations regarding necessary and sensible key change rates,
exhibiting the reflections that lead to the assumed requirements of a quantum key synchronization solution.
Section \ref{sec:rrp} contains the architecture of the presented solution and the subsequent Section
\ref{sec:impl} its implementation, while Section \ref{sec:ex:integration} describes its incorporation into the AIT QKD
software.
Descriptions of the setups and results of laboratory Experiments, showing the practical capabilities of this proof of
concept, form the Sections \ref{sec:ex:tp} through \ref{sec:disc}.
Section \ref{sec:con}, eventually, contains the conclusions drawn.

\section{Related Work}
\label{sec:Previous}
This work is aware of some previously developed methods to combine QKD with IPsec. All of them work in
conjunction with the IKE
\cite[pp.234-235]{EPT:2003}\cite[p.177-182]{QIKE:2008}\cite{MagiQ:2007}\cite[pp.5-9]{ID-nagayama-ipsecme-ipsec-with-qkd}\cite[p.21]{SLBCFGHJLM:2011}
or the underlying ISAKMP \cite[pp.6-8]{SGRG:2005} protocol. They introduce a supplement for QKD parameters or combine
IKE-derived and QKD-derived keys. Opposed to this, the presented work tries to use an approach omitting IKE and
following the pivotal idea that there is no need for that protocol to exchange keys, for that is the task of QKD. The key feed from QKD therefore provides the material for manual keying in this solution, all
that is left is to keep those keys synchronous. For this task, this paper proposes a more slender approach (see Section
\ref{sec:rrp}). Furthermore, some of the previous approaches
operate at a substantially lower speed than the key change  presented in this
thesis or use OTP limiting the data rate to the QKD key rate (currently around 52 kilobits per second) or simply suggest
applying QKD keys to IPsec without a mechanism for changing keys rapidly, effectively not lowering the number of data
bits per key bit.

\section{Key Change Rate Considerations}
\label{sec:keychange}
The strength of every cryptographic system relies on the key strength, the secrecy of the key and the effectiveness of
the used algorithms \cite[p.5]{NIST:2016}. As this solution relies on QKD, which
generates a secret and true random key \cite{PFUBLMPSKWHJZ:2004}, this means that more effective algorithms and more
key material are able to provide more cryptographic security. In this particular case, the used algorithms are already prescribed by the IPsec standard
\cite{IANA:2012}. Therefore, the security is mainly determined by the used key lengths, more precisely by the relation
between the amount of key material and the amount of data, which should be as much in favor of
the key material as possible - given the low key rate compared to the data rate, naturally the opposite is the case in
practice. This section aims on giving feasible upper and lower boundaries of key
change rates (or key periods $P_k$, respectively) and, thus, how much QKD key material should be used in order to save
precious quantum key material while maintaining a very high level of security.
The two main factors determining the key period in practice are the used algorithms (via their respective key
lengths - the longer the key, the more key bits are used in one key period) and the capabilities of QKD in generating
keys. The QKD solution of the Austrain Institute of Technology  has proven to provide a
quantum key rate $Q$ of up to 12,500 key bits per second at close distances, 3,300 key bits at around 25 kilometers and
550 key bits at around 50 kilometers distance \cite[p.9]{TPHFLQMHZ:2005}. As this paper presents a practical implementation (see Section
\ref{sec:impl}) in the form of a module for the AIT QKD software, the highest of these values should be the reference
key bandwidth for the key length and period considerations made in this section.

In order to fully utilize the possible QKD key rate and given the currently shortest
recommended key length, which is 128 bits (see below), an IPsec solution using quantum-derived keys should be able to
perform around 100 key changes per second ($\frac{12,500}{128}\approx97,65$), 50 for every communication direction (for IPsec
connection channels are in principle unidirectional and therefore independent from each other even if they belong to the
same bidirectional conversation). This corresponds to a key period $P_k$ of around 20 ms, as it is a function of the
Quantum key rate $Q$ and the algorithm's key length $k$. The period for a bidirectional IPsec link is
$P_K=({\frac{Q}{2k}})^{-1}$. At longer key lengths, this period becomes longer, for a single change cycle uses more key
material and, thus, less key changes are necessary to utilize the full incoming key stream, therefore this period
$P_{k_{min}}= 20 ms$ presents a feasible lower boundary for the key period.  
As stated above, the security of this system depends also on the data rate. Given a widespread data rate of 100 megabits
per second, a key period of 20 ms and 128 key bits means a ratio of 8000 data bits per key bit (or short dpk, for the
reader's convenience).

A landmark in this \textit{security ratio} is 1 dpk, as this rate would provide unconditional security
when applied with OTP. For the cipher and hash suites included in the IPsec protocol stack, there is no security proof
and therefore they are not unconditionally secure. However, applying an IPsec cipher (for instance AES) with an
appropriately fast key change and restricted data rate to achieve 1 dpk is the closest match inside standard IPsec,
especially when the block size equals the key size.  

To define an upper boundary (and therefore a minimum standard for the high security application of the presented
solution), a very unfavorable relation between data and key bits through a high-speed connection of 10 gigabits of
data is assumed. A recent attack on AES-192/256 uses $2^{69.2}$ computations with $2^{32}$ \textit{chosen plaintext}
\cite[p.1]{KJSHL:2013}. Because of the AES block size of 128 bits, this corresponds to $2^{32}*2^{7}=2^{39}$ data bits.
Although this attack is currently not feasible in practice, as it works only for seven out of 12/14 rounds and also has
unfeasible requirements to data storage on processing power for a cryptanalytic machine, it serves as a theoretical
fundament for this upper boundary. A bandwidth of 10 gigabits per second equals approximately 9.3 gibibits per second.
This is by the factor of 64 ($2^6$) smaller than the amount of data for the attack mentioned above, which means that it
requires 64 seconds to gather the necessary amount of data to (though only theoretically) conduct the attack. In
conclusion (with AES-192/256), the key should be changed at least every minute ($P_{k_{max}}=60 s$), while the maximum
allowed key period according to the IPsec standard lies at eight hours or 28,800 seconds \cite{RFC4308}.

For cryptographic algorithms operating with lower cipher block sizes ($\omega$), the
\textit{birthday bound} ($2^{\frac{\omega}{2}}$) is relevant. 
The birthday bound describes the number of brute force attempts to
 enforce a collision with a probability of 50 percent, such that different clear text
 messages render to the same cipher text \cite{kim2008birthday}.
With a block size of 64 (\textit{birthday bound} = $2^{32}$), the
example speed of 10 gigabit per second above would lower the secure key period to under half a second. Because of this
factor, using 64-bit ciphers is generally discouraged for the use with modern data rates\cite[pp.1-3]{McGrew:2012}
(although the present rapid rekeying approach is able to cope with this problem). 
Regarding key lengths, 128 bits are recommended beyond 2031 \cite[p.56]{NIST:2016} while key sizes of 256 bits provide
\textit{good protection} even against the use of Grover's algorithm in hypothetical quantum computers for this period
\cite[p.32]{ECRYPT:2012}.

\section{Rapid Rekeying Protocol}
\label{sec:rrp}
This section describes the \textit{rapid rekeying protocol}, the purpose of
which is to provide to IPsec peers with QKD-derived key material and keep these keys synchronous under the low-key-period conditions (down to $P_{k_{min}}= 20
ms$) stated in Section \ref{sec:keychange}.

\begin{figure}[htbp]
	\centering%
	\fbox{
		\includegraphics{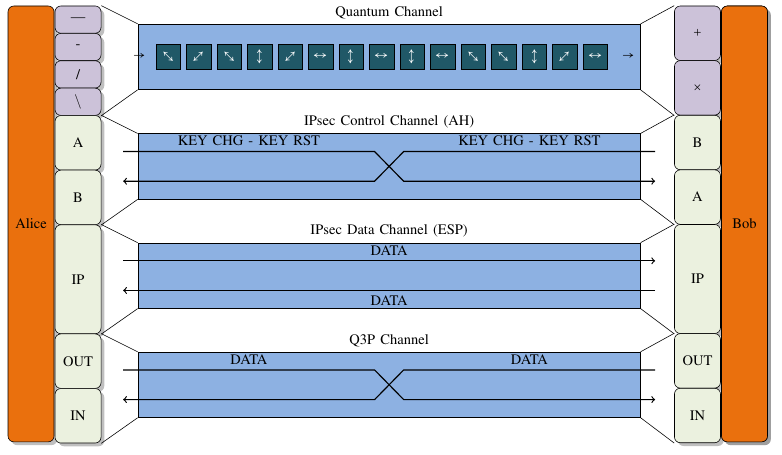}
	}
	\caption{Rapid Rekeying Channel Architecture}
	\label{img:Impl_Channels}
\end{figure}

This protocol pursues the approach that with QKD, there is no need for a classical key exchange (for instance
with IKE). Relevant connection parameters (like peer addresses) are available a priori (before the establishment of the
connection) in point-to-point connections, whereas keying material is provided by QKD, mostly obsoleting IKE.
Furthermore, IPsec only dictates an automatic key exchange, not specifically IKE
\cite[p.48]{RFC4301} and a protocol that only synchronizes QKD-derived keys (instead of exchanging keys)
is therefore deemed sufficient, yet compliant to the IPsec standard.
Consequently, it is an outspoken objective to create a slender and simple key synchronization protocol to increase
performance and reduce possible sources of error.  Another objective for key synchronization is robustness in terms of
resilience against suboptimal network environment conditions. The protocol described in this paper uses two channels
for encrypted communication:
an Authenticaton Header (AH)-authenticated control channel (amongst other tasks, signaling for key changes) and an
Encapsulating Security Payload (ESP)-encrypted data channel to transmit the protected data (see Figure
\ref{img:Impl_Channels}). The reason for the use of AH on the control channel is that it only contains non-secret
information, while its authenticity is crucial for the security and stability of the protocol. The necessary
\textit{security policies (SPs)} for the IPsec channels remain constant during the connection. There are four necessary
SPs, one data and one control SP for each direction.
The complete software solution will, delivered by the AIT QKD software, contain additionally the quantum channel for key
exchange and a \textit{Q3P} channel (see Section \ref{sec:ex:integration}), whereby the latter is another protocol that
provides OTP-encrypted QKD point-to-point links.

\medskip

The protocol itself follows, taking account of the unidirectional architecture of IPsec, a \textit{master/slave} paradigm.
Every peer assumes the master role for the connection in which the peer represents the sending part. When a key change
is due (for instance because of the expiration of the key period), the master sends an according message (key change
request) to the slave and the latter changes the key (as does the master). To compensate lost key change signals, every
key change message contains the \textit{security parameters index (SPI)} for the next-to-use key. 
The SPI is simply calculable for the peers through a salted hash whereby the salt and a initial seed value are
QKD-derived and each SPI is a hash of its predecessor plus salt, which makes it non-obvious to third parties. This level
of security is sufficient, for the SPI is a public value, included non-encrypted in every corresponding IPsec packet,
making it a subject rather to non-predictability than to secrecy. Also, using only a seed and salt from QKD, the
hashing method safes quantum keying material.
As all necessary IPsec parameters are available beforehand, as well as
the keys (through QKD), IPsec \textit{security associations (SAs)} may be pre-calculated and
established in advance (which are identified by unique SPIs). 
Permanently changing attributes during a conversation are only the SPI and the key, while all other parameters of an
SA (for instance peer addresses, services, protocols) remain constant. 
The master calculates these two in advance and queues them for future use. Only one SA is actually installed (aplied
to the kernel IPsec subsystem), for only one (per default, at least in Linux, the most recent) may be used to encrypt
data. The slave, on the other hand, operates differently. For it identifies the right key to use based on the SPI, it
may very well have multiple matching SAs installed. This makes key queuing expendable on the receiver side, while the
SPI queuing is used as an indexer for lost key change message detection. For reasons of data packets arriving out of
synchronization, SAs are not only installed beforehand, but also left in the system for some time on the receiver side,
allowing it to process packets encrypted with both an older or newer key than the current one. 

On every key change event, the master applies a new SA to the system (using the next
following SPI/key from the queues), prepares a new SPI/key pair (SPI generation as mentioned above and acquirement of a
new key from the QKD system) and deletes the deprecated data from both its queues and the IPsec subsystem.
The slave also acquires a new SPI/key pair (the same the sender acquires) but installs it directly as an SA and only
stores the SPI for indexing. It subsequently deletes the oldest SA from the system and SPI from the queue if the number
of installed SAs exceeds a configured limit.
To sum it up, on every key change event, the two peers conduct the following steps:
\begin{itemize}
\item the master acquires a new key and SPI and ads it to its queues
\item it sends a key change request to the slave
\item it fetches the oldest pair from the queue an installs it as a new SA, \textit{replacing} the current one
\item it deletes the deprecated pair from its queue
\item the slave receives the key change request and also acquires a new SPI/key pair (the same as the master)
\item it installs the pair as a new SA and the SPI into the indexing queue
\item it deletes the oldest SA from the system and oldest SPI from the queue
\item it sends a key change acknowledgement
\end{itemize}
This procedure keeps both of the installed SA types up to date.
For instance, 50 installed SAs for the slave resulting in 25 queued SPI/key pairs on the master, for the latter does
not need to store backward SAs. 
At the beginning, on every key change, SPI/key pair is acquired, while the already applied remain. When the
(configurable) working threshold is met, additionally the oldest SA or SPI/key pair is deleted, keeping the queue
sizes and number of installed SAs constant.

\begin{figure}[htbp]
	\centering%
	\fbox{
		\scalebox{0.9}{
			\includegraphics{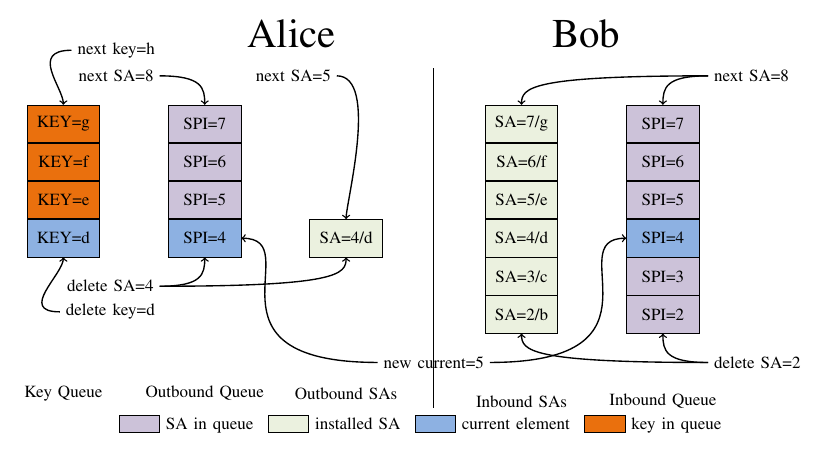}
		}
	}
	\caption{Key Change Process}
	\label{img:Impl_SA_Que} 
\end{figure}

Figure \ref{img:Impl_SA_Que} illustrates this process for a sender (\textit{Alice}) and a receiver (\textit{Bob}),
where the arrows show the changes in case of an induced key change. Naturally, as with SPs, there are four SA
types on a peer: one for data and control channels, each for sending (master) and receiving (slave). Each SA corresponds
to an SPI and key queue on the master's side and one SPI queue on the slave's side, respectively.
\medskip

\begin{figure}[htbp]
	\centering%
	\fbox{
		\scalebox{0.7}{
			\includegraphics{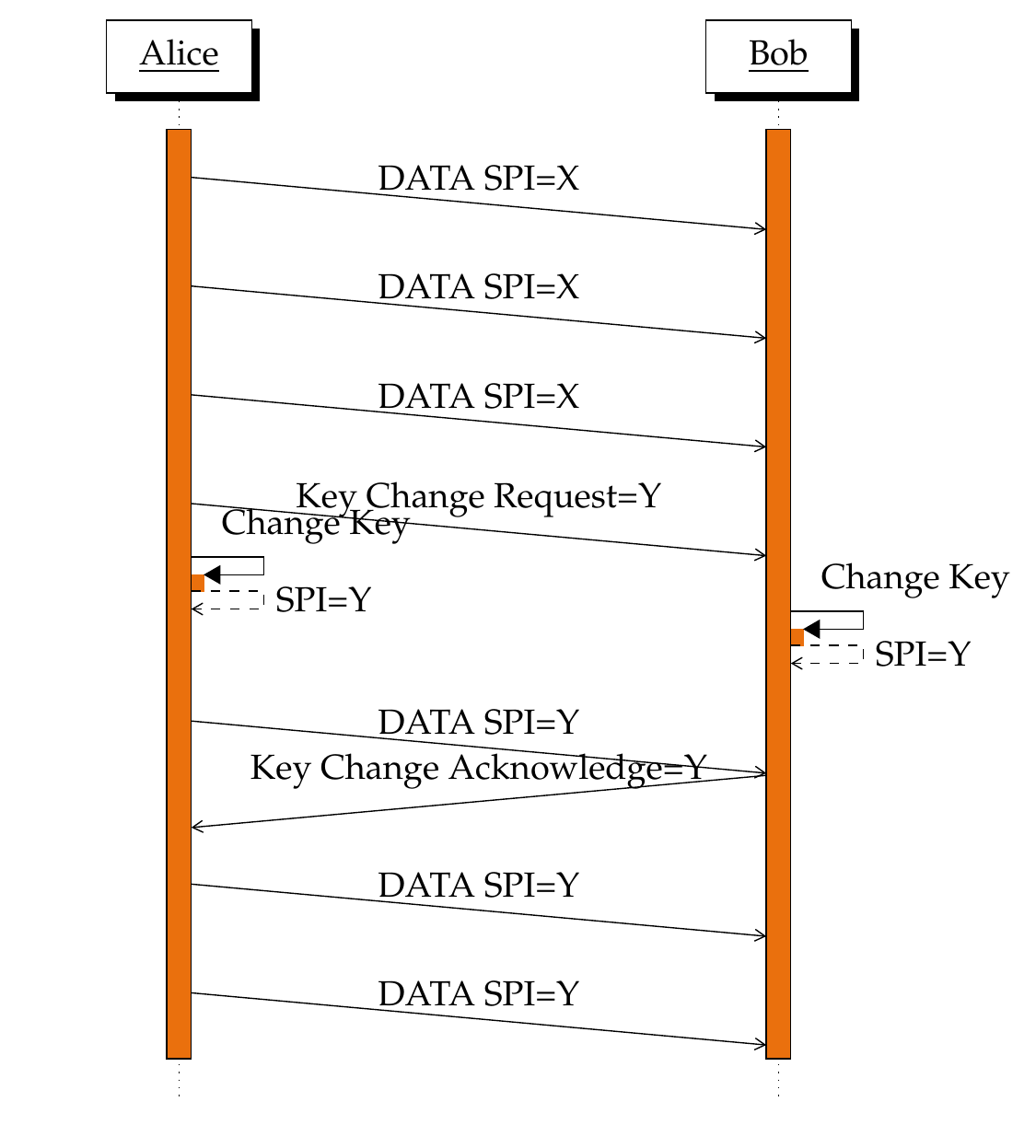}
		}
	}
	\caption{Key Change Message Flow}
	\label{img:Impl_Key_Change_Seq} 
\end{figure}

As the data stream is independent from control signaling, this calculation in advance prevents the destabilization of
the key synchronization in case of lost and too early or too late arriving key change messages. The buffer of
previously created SAs compensates desynchronization. For every receiver is able to calculate the according
SPIs beforehand, it may, by comparing a received SPI with an expected, detect and correct the discrepancy by calculating
the following SAs. Through this compensation process, there is neither need to interfere with the data communication nor
to even inform the sender of lost key change messages; the sender may unperturbedly continue with data and control
communications. This mechanisms make constant acknowledgements expendable and contribute thereby to a better protocol
performance through omission of the round trip times for the majority of the necessary control messages. Because of
this, acknowledgement messages (key change acknowledge) are still sent, but serve merely as a keepalive mechanism
instead of true acknowledgements (see Figure \ref{img:Impl_Key_Change_Seq}).
\medskip

In rare occasions, a key change message might be actually received, but the slave might not be able to apply the key
for some reason (for instance issues regarding the QKD system or the Kernel).
In this case, it reports the failure to the master with an appropriate message (key change fail).
In case too many control packets go missing (what the receiver is able to detect by SPI comparisons and the sender by
the absence of keepalive packets) or the key application fails, every peer is able to initiate a reset procedure (master
or slave reset). The actual threshold of allowed and compensated missing messages is a matter of configuration and
corresponds to the queue sizes for the SAs and therefore the ability of the system to compensate these losses. The
master does not need to report key change fails, for it is in control of the synchronization process and might just
initiate a reset if it is unable to apply its key. An additional occasion for a reset is the beginning of a conversation. At that point, the master starts the key
synchronization process with an initial reset. A reset consists of clearing and refilling all of the queues and
installed SAs.
For the same reason as for the data channel, the authentication key for
the control channel changes periodically. Due to the relatively low transmission rates on the control channel the key period is much longer (the software's default is
3 seconds) than on the data channel.
As, therefore, control channel key changes are comparatively rare and reset procedures should only occur in extreme
situations, both types implement a three way handshake. This is, on the one hand, because of the low impact on the
overall performance due to the rare occurrences, on the other hand due to higher impact of faulty packets. The control
channel, however, implements the same SA buffering method as the data channel (only with AH SAs, for the reasons
stated at the beginning of this section).

\section{Implementation}
\label{sec:impl}
\begin{figure}[htbp]
	\centering%
	\fbox{
		\includegraphics{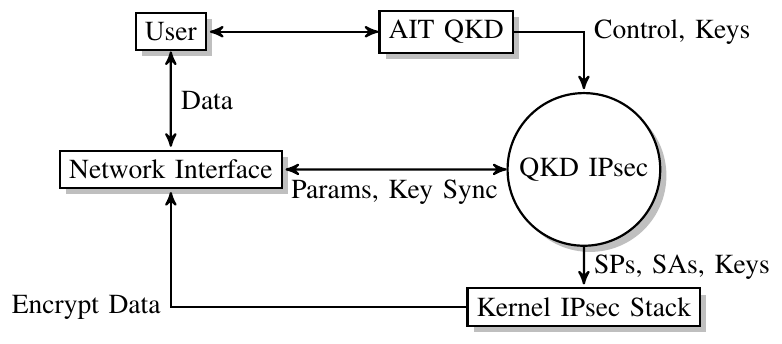}
	}
	\caption{QKDIPsec Systems Context}
	\label{img:Impl_Context} 
\end{figure}

The presented solution, called \textit{QKDIPsec}, consists of three parts (see also
Figure \ref{img:Impl_Context}):
\begin{itemize}
  \item key acquisition;
  \item key application;
  \item key synchronization;
\end{itemize}
\medskip

Each of this tasks has a corresponding submodule inside QKDIPsec, while the overall control lies within the
responsibility of the \textit{ConnectionManager} class, which provides the main outside interface and instantiates the
classes of said submodules using corresponding configuration. Also, all of these classes have corresponding
configuration classes using a factory method pattern \cite[p.134]{FRBS:2004} and according configuration classes,
decoupling program data and logic.
The first task (key acquisition) is the objective of an interface to the AIT QKD
software, the \textit{KeyManager}, which provides the quantum key material. In this proof of concept, this class generates
dummy key from a ring buffer, while it already has the according interfaces for the QKD software to serve as a class to
acquire quantum key material and provide it in an appropriate way to QKDIPsec. By now, only one function
implementation is missing on the QKD software side to fully integrate QKDIPsec into the QKD software.

The second part (\textit{KernelIPsecManager}) enters the acquired key directly into the Linux kernel, which
encrypts the data sent to and decrypts the data received from a peer.
Responsible for this part are a number of C++ classes, which control the SP and
SA databases (SPD and SAD) within the Kernel's IPsec subsystem via the Linux
\textit{Netlink} protocol. Therefore, this solution uses the derived class \textit{NetlinkIPsecManager}, but leaves the
option to use other methods for kernel access as well. The reason for using Netlink to communicate with the kernel
is that it was found the most intuitive of the available methods and that it is also able to handle not only the
IPsec subsystem but a broad span of network functions in Linux. Furthermore, using a direct
kernel API, as opposed to other IPsec implementations, omits middleware, both enhancing performance as well as
eliminating potential source of error.
Also using Netlink functions, this part governs the tunnel interfaces and routing table
entries necessary for the communication via the classes \textit{KernelNetworkManager} and \textit{NetlinkNetworkManager}
as well.

Netlink is a socket-oriented protocol and allows therefore the use of well-known functions from network programming.
The difference to the latter is that instead of network peers, communication runs within the system as
\textit{inter-process communication (IPC)}, through which also the kernel (via process ID zero) is addressable. Due to its
network-oriented nature, a packet structure is used instead of function calls via parameters. This means that commands
to the kernel (for instance to add a new SA) needs to be memory-aligned in the
according packet structure and subsequently send to the kernel via a Netlink socket.
A downside of Netlink during implementation was the complicated nature and weak documentation of its IPsec manipulation
part (\textit{NETLINK\_XFRM}). While the Netlink protocol itself is present in every message in the form of its uniform
header, the \textit{NETLINK\_XFRM} parts use a different structure plus individual extra payload attributes for every
type of message (add and delete messages for both SAs and SPs), making the according class hierarchy rather inflated. Also, the
solution uses the \textit{NETLINK\_ROUTE} protocol to add and delete both IP interface addresses and network routes.

To take this into account, the QKDIPsec implementation uses a set of Netlink message classes, deriving from the common
base class \textit{NetlinkMessage}. This class contains the common Netlink header. Each message type for IPsec and
network function configuration is further a child class, containing the exact data fields necessary for Netlink. Due to
the separation of code and data segments in C++, the class functions do not interfere with the netlink data fields and
therefore its alignment \cite[pp.142-143]{vdLinden:1994}. This means that the class hierarchy takes care of the memory
alignment necessary for the Netlink protocol.
As stated above, the structure for \textit{NETLINK\_XFRM} messages is rather heterogenous, basically requiring every
message type to be assembled directly in the class, except for the Netlink header. The messages of the
\textit{NETLINK\_ROUTE} protocol, on the other hand, are more structured, allowing it to introduce intermediate classes
for routing table and interface addresses messages.

The key synchronization, eventually, is the main task of the \textit{Rapid
Rekeying Protocol}. As this is the very core of the solution, its
implementation resides directly inside the connection manager. While it uses the classes mentioned above to acquire and
apply the QKD keys in the manner discussed in Section \ref{sec:rrp}, it handles
the key synchronization using sender and receiver threads (representing the master and slave parts, respectively), as well as a class for key synchronization 
messages. Within this class, also the described lost message compensation and reset, as well as initialization
and clean-up procedures are implemented.
The reset procedure may also include some re-initialization process for the QKD system, triggered via the
\textit{KeyManager}. This class also sets the clocking for the key changes, which is dynamically adjustable during
runtime.

\section{Integration}
\label{sec:ex:integration}

QKDIPsec has been integrated into the current AIT QKD R10 Software Suite V9.9999.7\cite{Maur:2015}. 
This Open Source software contains a full featured QKD post processing environment containing BB84
sifting, error correction, privacy amplification and other steps necessary. The final stage of an 
AIT QKD post processing pipeline is a QKD key store, realized as \textit{Q3P} link.

\begin{figure}[htbp]
	\centering%
	\fbox{
		\includegraphics[width=0.455\textwidth]{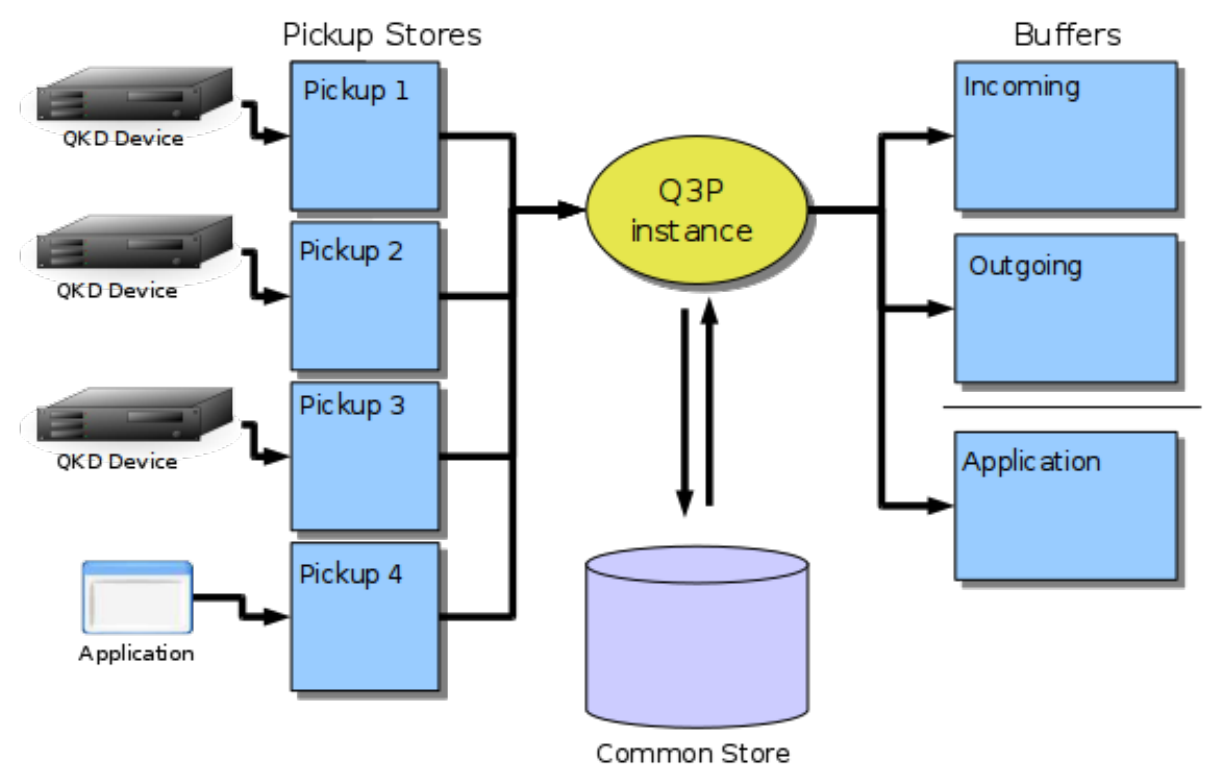}
	}
	\caption{Q3P Key Store Model}
	\label{img:Q3PKeyStores} 
\end{figure}

The central task of \textit{Q3P} is to keep the key material derived from quantum key distribution 
in synchronization on both ends of a point-to-point link. It does this by managing several buffers
as depicted in Figure \ref{img:Q3PKeyStores}.
\medskip

\begin{itemize}

\item \textit{A Pickup Store}: Before a key can be used, Q3P has to verify, that a particular key is
present on the other side of the connection. Reasons a key may not be present at the
same point of time in a peer's key store are found in the highly asynchrony and distributed
manner key material is inserted on both machines. Therefore, those key blocks are
treated as a collection of potentially usable keys and are stored in a Pickup Store directly related
to a certain QKD post processing pipeline. Hence, a single Q3P link can maintain multiple concurrent QKD 
post processing lines to boost throughput. Also Q3P does not know if a concrete QKD hardware device is
pushing keys into the Pickup Store or an application, which might have derived shared secret keys by other
means of deployment.

\item \textit{A Common Store}: Once the presence of the key material has been verified on both sides the key
is transfered to the \textit{Common Store} on disk. This is the only persistent data storage of key material
within Q3P. However, keys placed in the Common Store are not bound to any dedicated usage.

\item \textit{An Outgoing Buffer}: Once key material is present in the Common Store, Q3P moves chunks of key
material to an \textit{Outgoing Buffer}. Keys residing in this buffer are used to establish an information- 
theoretically secure channel for encryption and authentication for outgoing messages. Note that, due to the nature
of information-theoretically secure ciphers (such as the Vernam cipher),
 encryption combined with authentication key consumption for single messages is at a
minimum as large as the length of the message sent \cite[p.15]{Vernam:1926}. Also, keys that are used for messaging
are removed from the buffer and destroyed.
 
\item \textit{An Incoming Buffer}: For incoming messages each Q3P endpoint mirrors the Outgoing Buffer of its 
peer as its local \textit{Incoming Buffer}. The keys for authenticity checks of received messages as well as for
decryption are picked from this buffer. 
 
\item \textit{An Application Buffer}: On behalf the Incoming and the Outgoing Buffers Q3P established yet a third
Buffer: the \textit{Application Buffer}. Key material moved from the Common Store to this buffer in memory is
dedicated for use by any application utilizing Q3P.
  
\end{itemize}

The rationale for having separate buffers for outgoing messages and one for incoming is based on potential 
race conditions when doing heavy communication in both directions. Suppose both Q3P nodes do heavy interaction 
in streaming messages in both directions, then without such separation the situation, in which both key 
stores utilize the very same key for different messages is most likely. Q3P also introduces a master/slave
role model on key dedication: one partner in the communication acts as master, which is responsible for 
assigning key material from the Common Store to one of the three buffers. The slave on the other side requests
such assignments on demand.

The filling of the Outgoing and Incoming Buffers take precedence before the Application Buffer. Only if both buffers
used for direct information theoretic communication do share a minimum threshold of key material the Application
buffer is filled with keys from the Common Store.

The proposed protocol uses the established information theoretic secured channel provided by Q3P by means of the
Outgoing and Incoming Buffer inside Rapid Rekeying. Key material from the Application Buffer is used to create
the protocols SPI and SAs. As key material is directed to the Outgoing and Incoming Buffers first, this results 
in ``slow start'' of an IPSec enabled connection.

Although the protocol runs inside the process space of a single Q3P instance, from a software engineering point
of view the protocol's key withdrawal of the Application Buffer bears no difference to any other application
using the same buffer.

\section{Throughput Experiments}
\label{sec:ex:tp}
\begin{figure}[htbp]
	\centering%
	\fbox{
		\includegraphics[width=0.455\textwidth]{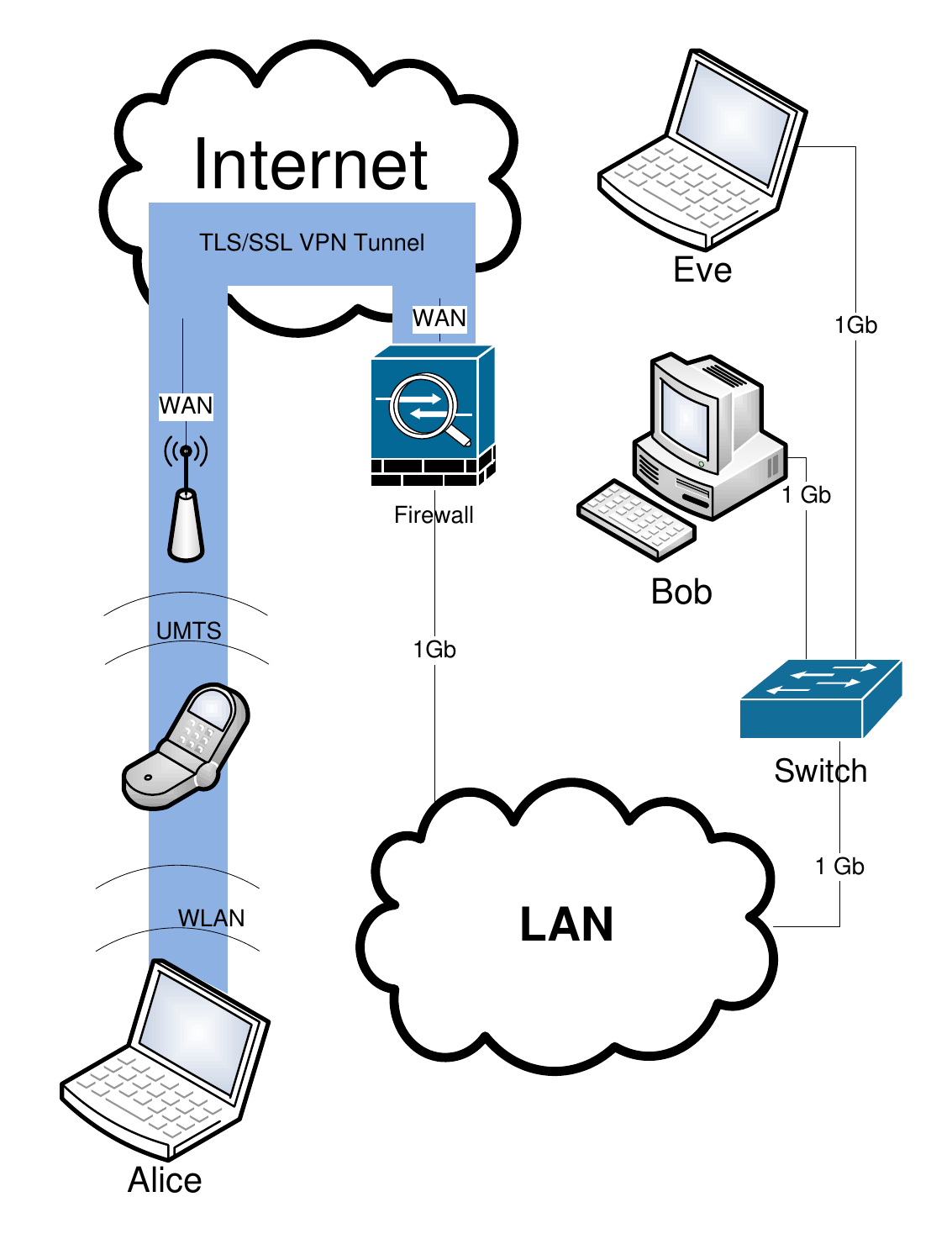}
	}
	\caption{WAN Test Setup}
	\label{img:WAN} 
\end{figure}

The protocol design of the described solution aims on the one hand on speed and
flexibility and on the other hand on fault tolerance, hence the
architecture is as simple and lightweight as possible (including abandoning the
IKE protocol). Due to this, very high IPsec key change rates can be achieved,
even under harsh conditions. The solution was implemented in software using C++
and tested on two to five year-old Linux computers (Alice and Bob), both in a
gigabit Local Area Network (LAN) and a UMTS-Wide Area Network (WAN) environment (the latter further aggravated by
combining it with WLAN and an additional TLS-based VPN tunnel - see Figure \ref{img:WAN}) by means of
data transfer time measurement and ping tests, as well as validation of the actual key changes
by a Wireshark network sniffer (Eve).
\medskip

Table \ref{fig:table:Perf} shows the results in seconds (four trials each, separated by slashes)
of data transmission and in percent on ping tests within the mentioned LAN and WAN environments with various
configurations:
unencrypted, standard IPsec and QKDIPsec with different encryption algorithms, the latter also with different key
periods.
In these tests, both data transfer and ping were initiated by one peer (\textit{Alice}). While the ping test was
continuous, the data transfer consisted each of one data transfer from \textit{Alice} to \textit{Bob} and vice versa. The test
file used on the LAN was a video file of 69.533.696 bytes size, while the WAN file was also a video, but only 1.813.904
bytes big.
In both cases, key periods of 25 ms and less could be achieved, maintaining a stable data connection. This, using the
recommended key length of 256 bit, surpasses the goal of 12,500 key bits per
second (the currently maximal quantum key distribution rate under ideal
circumstances), even though (deliberately) legacy equipment and a
less-than-ideal network environment was used. Comparison of the performance shows a (expectable) higher data transfer
period of QKDIPsec and unencrypted traffic, but no significant difference to
traditional IPsec. Only the packet losses on a simultaneously running ping
test were a few percentage points higher (mainly in the WAN environment). 
\captionsetup{font={footnotesize,sc},justification=centering,labelsep=period}%
 \begin{table}[htbp]
 	\caption{Performance Test Results}
 	\label{fig:table:Perf}
 	\centering%
	\begin{tabularx}{0.45\textwidth}{|X| c| c|c|}
	   \hline
	   \multicolumn{4}{|c|}{\textbf{LAN}}\\
	    \hline                         
	   	Setting &A$\rightarrow$B&B$\rightarrow$A&Ping\\
		\hline			  
		unencrypted	 &6/6/7/6&7/9/7/8&100\%\\
		\hline
		AES-256 CCM&&&\\
		standard IPsec& 14/14/16/15&17/18/26/18 &100\%\\
		50 ms  &8/10/8/9 &14/16/16/16&100\%\\
		25 ms  &10/9/8/8&14/15/17/16 &100\%\\
		20 ms  &9/9/9/9 &11/16/17/12 &100\%\\
		\hline
		AES-256 CBC&&&\\
		20 ms &9/7/7&11/13/17&100\%\\
		\hline
		Blowfish-448&&&\\
		20 ms &14/9/7&15/13/14&99\%\\
		\hline
	   \multicolumn{4}{|c|}{\textbf{WAN}}\\
	    \hline                         
	   	Setting &A$\rightarrow$B&B$\rightarrow$A&Ping\\
		\hline				  
		unencrypted	 &10/10/10/10&9/7/6/7&99\%\\
		\hline
		AES-256 CCM&&&\\
		standard IPsec& 11/11/11/11&11/5/6/5&99\%\\
		50 ms  &14/10/11/13&6/5/5/5&95\%\\
		25 ms  &10/11/10/10&6/7/6/7&94\%\\
		20 ms  &12/11/13/10&9/5/6/6&98\%\\
		\hline
		AES-256 CBC&&&\\
		20 ms &10/11/11&9/7/8&100\%\\
		\hline  	
	\end{tabularx}
\end{table}
\captionsetup{font={footnotesize,rm},justification=centering,labelsep=period}%

To verify the key changes, a network sniffer, Eve, was keeping track of the actual SPI changes of the packets
transmitted between Alice and Bob. Table \ref{fig:table:Sniff} shows
a random sample of key change periods in milliseconds during the above mentioned LAN 20ms
AES-256-CCM test. Within this table, the first column shows the key change times for data (ESP) packets from Alice to
Bob while the second shows the opposite direction.
As the recorded data contains one file copy from Alice to Bob (in the first half of the record) and one vice versa (in the second half),
one randomly chosen sample of five consecutive key changes for each direction and from each half is chosen. 
This form of sample choosing from different phases and directions of the communication session and averaging them
compensates inaccuracies, induced by the pause between key change and respective next following packet, which become
greater the less traffic is sent. As the receiver only acknowledges received data and, therefore, sends significantly
less packets, the vagueness of the non-averaged results is greater when receiving.
The total average of all four of these averaged values is $0.020495$ ms, which is approximately $2.5\%$ above 20
ms per key change. This may be explained by the send
and receive overhead for processing the key change messages, for the period
determines only the sleeping duration of a sender thread.  

Because of the lower amount of traffic (due to the lower speed) and higher latency such exact time readings are not
possible in the WAN environment. Therefore, the measurement method was changed to averaging a sample set of 20 key change periods,
using the same random choosing as above. With approximately $0.2475$, the total averaged result lies significantly
higher (approximately $19\%$) than the one of the LAN setting. One possible explanation for this behavior is the latency
in this environment.
\begin{table}[htbp]
	\caption{Network Sniffing Results}
	\centering%
	\label{fig:table:Sniff}
	\begin{tabular}{r| c| c|c| c|}
	   \hhline{~|-|-|-|-|}
	    &\multicolumn{2}{c|}{A$\rightarrow$B}&\multicolumn{2}{c|}{B$\rightarrow$A}\\
	   \hhline{~|-|-|-|-|}                         
	   		 &1st&2nd&1st&2nd\\
		\hhline{~|=|=|=|=|}			  
		LAN				&0.0220&0.0216&0.0208&0.0203\\
						&0.0187&0.0204&0.0197&0.0235\\
						&0.0145&0.0216&0.0203&0.0176\\
						&0.0195&0.0243&0.0204&0.0197\\
						&0.0225&0.0180&0.0207&0.0238\\
		\hhline{~|-|-|-|-|}
		\O				&0.0194&0.0212&0.0204&0.0210\\
		\hhline{~|=|=|=|=|}
		WAN \begin{small}$\sum 20$\end{small}	&0.5201&0.4899&0.4302&0.5397\\
		\O				&0.0260&0.0245&0.0215&0.0270\\
		\hhline{~|-|-|-|-|}
	\end{tabular}	
\end{table}

Additionally, the recovery behavior was tested by letting the master deliberately omit key change
notifications through manipulating the sending routine, while again running ping tests and
file copies. Omitting single key change messages (and, thus, testing the recovery mechanism) yield in no measurable
impact on the connection (along with 100\% of successful pings). Also, by the same method of omitting key change
requests, but this time surpassing the recovery queue size, the reset procedure was tested. The queue size was set to
50 and \textit{Alice} was programmed to omit 50 sending key change messages after 200 sent ones. Expectedly, \textit{Bob}
initiated a reset procedure during the hiatus, resulting in a cycle of 200 key changes and a subsequent reset. Despite
these permanent reset-induced interruptions, bidirectional ping tests only yielded insignificant losses (99.74\% from
\textit{Alice} to \textit{Bob} and 99.36\% vice versa). Furthermore, a file copy
in both directions was still possible.

Further, to test the endurance of the solution, one experiment was conducted to show the capability of maintaining the
connection over a longer period of time. It was performed with an earlier development version of QKDIPsec and ran in LAN
environment over around 16 hours. It consisted of a running ping test on a 50 ms \textit{Blowfish} configuration without
control channel key changes. Of 56179 pings returned 56164 resulting in a return rate of approximately $99.97\%$. This
test was also conducted in WAN environment, but (due to both tests ran overnight) an automated network connection reset
after around eight hours prevented meaningful results.

The last test was actively severing the network connection. Pulling the plug on one side resulted in a connection loss
that was only recoverable by executing the connection setup routine. This normally does not occur automatically in
QKDIPsec but can be induced by the calling function (ordinarily the AIT QKD software). The cause for this
behavior is that a shut down (or connectionless) interface loses its additional IP addresses and therefore the tunnel address for
the data channel. This problem might be circumvented by implementing an own virtual interface in the future.
When servering the connection along the path (thus leaving the peer interfaces intact) the solution automatically
recovered (loosing only traffic during the servered phase) when reconnected timely or entered the reset procedures
(reset trial and function suspension on time) on disruption spanning over more than the timeout period, according to
protocol.

\section{QKDIPsec in a Simulation}
\label{sec:disc}

In order to investigate the impact of the time interval between key change notifications on the overall performance and on the underlying data 
transmission, we implemented the~\textit{Rapid Rekeying Protocol} in OMNeT++~\cite{varga2001omnet++} using the INET framework. Besides IPSec and 
the~\textit{Rapid Rekeying Protocol} we implemented an UDP application that sends a certain amount of data to its
counterpart using IPSec. We built an evaluation setup with two communicating hosts, and introduced delay and packet
drops to the setup.
The~\textit{Rapid Rekeying Protocol} allows to vary the following variables: number of 
(simultaneous) installed SAs, and the interval between sending a key change request. For now, we assume that the keys
can be provided with an infinite rate, thus idealizing the generation of the key material. Table~\ref{fig:table:paramsettings} provides the different parameter 
settings used for the simulation. For each combination of the parameters (64 in total) we conducted 30 runs. For the simulation we assumed a sufficiently 
large QKD key rate (such that none of the applications has to wait for new key material). In the following we report the averages of these runs and their 95\% 
confidence interval (CI) for some selecting parameter settings.

\begin{table}[htbp]
	\caption{Parameter Settings for the Simulation.}
	\label{fig:table:paramsettings}
	\centering%
	  \begin{tabular}{| c | c | }
        \hline
        Parameter &  Values \\ \hline
        \hline                      
        Installed SAs & 5, 15, 40, 70 \\ \hline
        Key Change Interval (ms) & 25, 50, 100, 200 \\hline
        UDP Data Traffic (Mbps) & 1, 1.5, 1.7, 1.9 \\ \hline
        Simulation Time (s) & 600 \\ \hline
        Channel Delay (ms) & $X \sim U(5, 25)$ \\ \hline
        Channel Data Rate (Mbps) & 2 \\ \hline
        Packet Drop Probability & min($X \sim U(0, 1)$ , 0.05) \\ \hline
    \end{tabular}
\end{table}

\begin{figure}[htbp]
	\centering%
	\fbox{
		\includegraphics[width=0.455\textwidth]{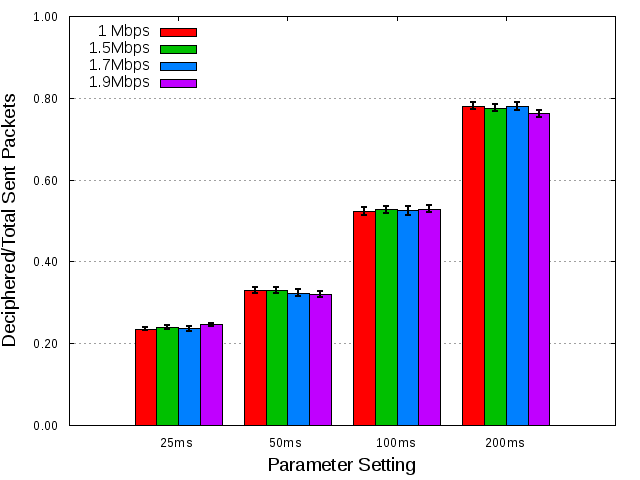}
	}
	\caption{Packets deciphered relative to the total amount of sent packets for the given data rates with a 
	maximum number of 5 installed SAs for different re-keying intervals, respectively.}
	\label{img:deciphered:10} 
\end{figure}

\begin{figure}[htbp]
	\centering%
	\fbox{
		\includegraphics[width=0.455\textwidth]{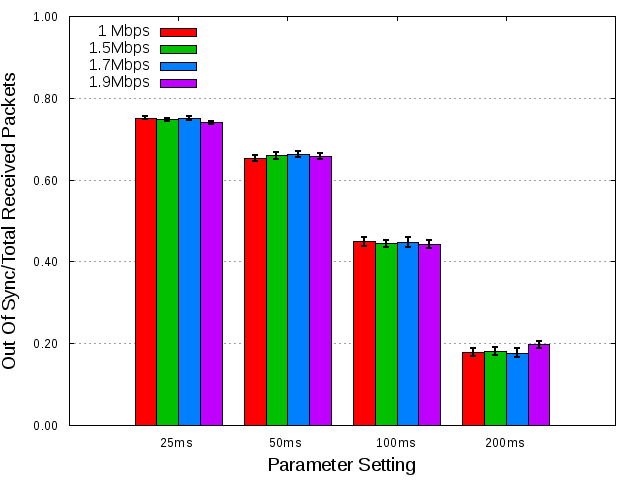}
	}
	\caption{Packets out of synchronization relative to the total amount of received packets for the given data rates with a 
	maximum number of 5 installed SAs for different re-keying intervals, respectively.}
	\label{img:outofsync:10} 
\end{figure}

Figure~\ref{img:deciphered:10} depicts the average of deciphered packets with 95\% CIs for a maximum of 5 installed SAs at the receiving client. With an 
increase in the re-keying interval the receiver is able to decipher approximately 80\% of all data packets. This is valid for the tested data rates. 
Although, reaching the theoretical channel data rate of 2~Mbps decreases the number of deciphered packets due to the fact that packets are dropped by full 
queues. Figure~\ref{img:outofsync:10} depicts the average of out of synchronization packet with 95\% CIs relative to the total amount of received packets using 
the same 
parameter settings as for Figure~\ref{img:deciphered:10}. It is evident that with a lower re-keying intervals the amount of non-decipherable and out of 
synchronization 
packets increases. However, selecting larger re-keying intervals increases the probability that a man in the middle attacks will be successful. Therefore, a 
tradeoff between data rate and the desired security level has to be found. Although, we have to consider that some packets are dropped because of the chosen 
packet drop 
probability (cf. Table~\ref{fig:table:paramsettings}). 

\begin{figure}[htbp]
	\centering%
	\fbox{
		\includegraphics[width=0.455\textwidth]{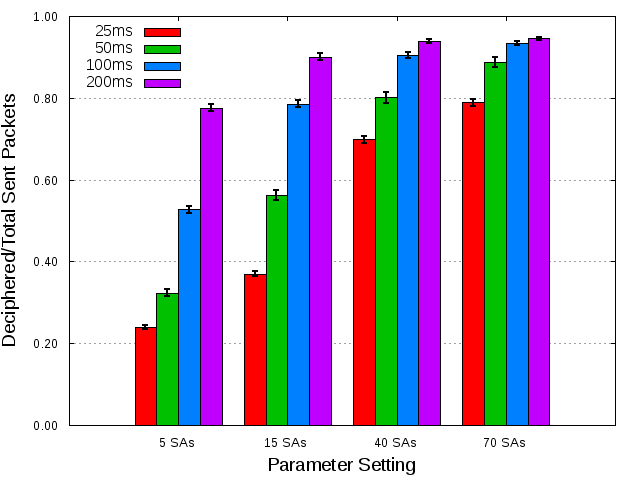}
	}
	\caption{Packets deciphered relative to the total amount of sent packets for a data rate of 1.5~Mbps.}
	\label{img:deciphered:80} 
\end{figure}

\begin{figure}[htbp]
	\centering%
	\fbox{
		\includegraphics[width=0.455\textwidth]{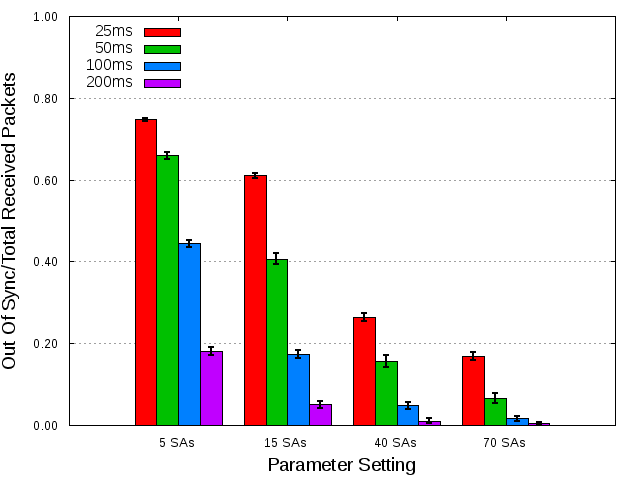}
	}
	\caption{Packets out of synchronization relative to the total amount of received packets for a data rate of 1.5~Mbps.}
	\label{img:outofsync:80} 
\end{figure}

Figures~\ref{img:deciphered:80} and~\ref{img:outofsync:80} depict the relative amount of deciphered and out of synchronization packets for a data rate of with 
95\% CIs for 
a data rate of 1.5~Mbps. Increasing the the number of simultaneous installed SAs, the 
probability of encountering out of synchronization packets decreases. Nonetheless, one observes the same behavior as for Figures~\ref{img:deciphered:10} 
and~\ref{img:outofsync:10}. Assuming a re-keying interval of 100~ms, a data rate of 1.5~Mbps and a maximum of 15 installed SAs, using QKDIPsec we are able 
to achieve an effective data rate of approx. 1.1~Mbps on average. If a re-keying interval of 200~ms is acceptable, we
are able to achieve an effective data rate of approximately 1.35~Mbps on average. 
However, it remains the ultimate goal to derive a model by means of $\varepsilon$-security, which provides a trade-off 
between security and the effective data rate. We devote this to future work.

\section{Conclusion}
\label{sec:con}

These proof of concept tests show that using IPsec with appropriate key management is able to overcome the bandwidth
restrictions of QKD, even when operating the data channels in less-than-ideal conditions.
This, however, comes with the cost of having to reuse the key more than once. Therefore, this paper discussed sensible
boundaries of key usage to maintain an acceptable level of security (see Section \ref{sec:keychange}).
Furthermore, this paper presents an approach to provide QKD-secured links with
high speeds meeting the bounds discussed in Section \ref{sec:keychange}, including a suitable performant and
fault-tolerant key synchronization protocol (the \textit{rapid rekeying protocol}) and a corresponding software solution
running under Linux (\textit{QKDIPsec}), integrated into the AIT QKD software. Furthermore, this proof of concept was
thouroughly tested both on x86 system architectures and in a simulated machine environments. These tests
showed the operability of the principal architecture design as well as possible snares regarding its implementation.
During these tests, it became obvious that more installed SAs increase the rate of sucessfully deciphered packets,
especially in lower key period settings.

Despite promising test results, there is room for improvement to transform the presented proof of concept module into a
fully productive and integrated part of the AIT QKD software. Firstly, there are still obstructions to tackle regarding
the integration; the methods for key capturing from the Q3P Application Buffer have to be elaborated and optimized.
Secondly, further tests are needed to determine the optimal choice of networking mechanisms. For instance, the implications of switching
from TCP to UDP as a transport layer protocol for QKDIPsec have to be examined. Thirdly, some procedures have to be
introduced, which automate the reset process in case of hardware connection losses and resets, eliminating the need to
restart the system manually. Fourthly, to ease its setup, the solution needs the ability to use virtual interfaces as
tunnel endpoints (currently it only supports virtual addresses). Fiftly, while the current version of QKDIPsec already
supports on-the-fly adjustments of the key period, the solution should be able to provide interfaces to automatically
align this key period to a desired rate of data bits per key bit (dpk). This makes it necessary to provide means to
measure the actual data rate running over the data channel and comparing them to the key effective key change rate
(consisting of key period and key length). Furthermore, it is desirable to  derive a model by means of
$\varepsilon$-security, to achieve a trade-off between the data rate and the security of this solution.  



%
%
%

\bibliographystyle{IEEEtran}
\bibliography{literature}

\begin{thebibliography}{10}
\providecommand{\url}[1]{#1}
\csname url@samestyle\endcsname
\providecommand{\newblock}{\relax}
\providecommand{\bibinfo}[2]{#2}
\providecommand{\BIBentrySTDinterwordspacing}{\spaceskip=0pt\relax}
\providecommand{\BIBentryALTinterwordstretchfactor}{4}
\providecommand{\BIBentryALTinterwordspacing}{\spaceskip=\fontdimen2\font plus
\BIBentryALTinterwordstretchfactor\fontdimen3\font minus
  \fontdimen4\font\relax}
\providecommand{\BIBforeignlanguage}[2]{{%
\expandafter\ifx\csname l@#1\endcsname\relax
\typeout{** WARNING: IEEEtran.bst: No hyphenation pattern has been}%
\typeout{** loaded for the language `#1'. Using the pattern for}%
\typeout{** the default language instead.}%
\else
\language=\csname l@#1\endcsname
\fi
#2}}
\providecommand{\BIBdecl}{\relax}
\BIBdecl

\bibitem{MM:2015}
S.~Marksteiner and O.~Maurhart, ``{A Protocol for Synchronizing Quantum-Derived
  Keys in IPsec and its Implementation},'' in \emph{{ICQNM 2015, The Ninth
  International Conference on Quantum, Nano and Micro Technologies.}},
  V.~Privman and V.~Ovchinnikov, Eds.\hskip 1em plus 0.5em minus 0.4em\relax
  Venice: IARIA, 2015, pp. 35--40.

\bibitem{ZBGR:1998}
H.~{Zbinden}, H.~{Bechmann-Pasquinucci}, N.~{Gisin}, and G.~{Ribordy},
  ``\BIBforeignlanguage{English}{{Quantum cryptography}},''
  \emph{\BIBforeignlanguage{English}{{Applied Physics B}}}, vol.~67, no.~6, pp.
  743--748, 1998.

\bibitem{NC:2000}
M.~A. {Nielsen} and I.~L. {Chuang}, \emph{\BIBforeignlanguage{English}{{Quantum
  Computation and Quantum Information}}}, ser. {Lecture Notes in
  Physics}.\hskip 1em plus 0.5em minus 0.4em\relax Cambridge: Cambridge
  University Press, 2000.

\bibitem{Shannon:1949}
C.~E. {Shannon}, ``{Communication Theory of Secrecy Systems},'' \emph{The Bell
  System Technical Journal}, vol.~28, pp. 656--715, October 1949.

\bibitem{WHHLPZ:2015}
C.~Wang, D.~Huang, P.~Huang, D.~Lin, J.~Peng, and G.~Zeng, ``{25 {MHz} clock
  continuous-variable quantum key distribution system over 50 km fiber
  channel},'' \emph{Scientific Reports}, vol.~5, p. 14607, 2015.

\bibitem{TPHFLQMHZ:2005}
A.~Treiber, A.~Poppe, M.~Hentschel, D.~Ferrini, T.~Lor{\"u}nser, E.~Querasser,
  T.~Matyus, H.~H{\"u}bel, and A.~Zeilinger, ``{A fully automated
  entaglement-based quantum cryptography system for telecom fiber networks},''
  \emph{New Journal of Physics}, no.~11, p. 045013, April 2009.

\bibitem{SK:2010}
P.~{Schartner} and C.~{Kollmitzer},
  ``\BIBforeignlanguage{English}{{Quantum-Cryptographic Networks from a
  Prototype to the Citizen}},'' in \emph{\BIBforeignlanguage{English}{{Applied
  Quantum Cryptography}}}, ser. {Lecture Notes in Physics}, C.~{Kollmitzer} and
  M.~{Pivk}, Eds.\hskip 1em plus 0.5em minus 0.4em\relax Berlin, Heidelberg:
  Springer, 2010, vol. 797, pp. 173--184.

\bibitem{XCWYZLZZLLHG:2009}
F.~Xu, W.~Chen, S.~Wang, Z.~Yin, Y.~Zhang, Y.~Liu, Z.~Zhou, Y.~Zhao, H.~Li,
  D.~Liu, Z.~Han, and G.~Guo, ``\BIBforeignlanguage{English}{{Field experiment
  on a robust hierarchical metropolitan quantum cryptography network}},''
  \emph{\BIBforeignlanguage{English}{Chinese Science Bulletin}}, vol.~54,
  no.~17, pp. 2991--2997, 2009.

\bibitem{RFC4301}
S.~{Kent} and K.~{Seo}, ``{Security Architecture for the Internet Protocol},''
  Internet Requests for Comments, {Internet Engineering Task Force}, {RFC}
  4301, 2005.

\bibitem{EPT:2003}
C.~Elliott, D.~Pearson, and G.~Troxel, ``Quantum cryptography in practice,'' in
  \emph{Proceedings of the 2003 conference on Applications, technologies,
  architectures, and protocols for computer communications}.\hskip 1em plus
  0.5em minus 0.4em\relax ACM, 2003, pp. 227--238.

\bibitem{QIKE:2008}
A.~Neppach, C.~Pfaffel-Janser, I.~Wimberger, T.~Lor{\"u}nser, M.~Meyenburg,
  A.~Szekely, and J.~Wolkerstorfer, ``Key management of quantum generated keys
  in ipsec.'' in \emph{Proceedings of SECCRYPT 2008}.\hskip 1em plus 0.5em
  minus 0.4em\relax INSTICC Press, 2008, pp. 177--183.

\bibitem{MagiQ:2007}
\BIBentryALTinterwordspacing
{MagiQ Technologies}, ``{MAGIQ QPN 8505 Security Gateway},'' 2007, retrieved at
  November 11, 2016. [Online]. Available:
  \url{http://www.magiqtech.com/Products\_files/8505\_Data\_Sheet.pdf}
\BIBentrySTDinterwordspacing

\bibitem{ID-nagayama-ipsecme-ipsec-with-qkd}
S.~{Nagayama} and R.~{Van Meter}, ``{Internet-Draft: IKE for IPsec with QKD},''
  2009, draft-nagayama-ipsecme-ipsec-with-qkd-00, expired work.

\bibitem{SLBCFGHJLM:2011}
D.~Stucki, M.~Legr{\'e}, F.~Buntschu, B.~Clausen, N.~Felber, N.~Gisin,
  L.~Henzen, P.~Junod, G.~Litzistorf, P.~Monbaron \emph{et~al.}, ``Long-term
  performance of the swissquantum quantum key distribution network in a field
  environment,'' \emph{New Journal of Physics}, vol.~13, no.~12, p. 123001,
  2011.

\bibitem{SGRG:2005}
M.~Sfaxi, S.~Ghernaouti-H{\'e}lie, G.~Ribordy, and O.~Gay, ``Using quantum key
  distribution within ipsec to secure man communications,'' in
  \emph{Proceedings of Metropolitan Area Networks (MAN2005)}, 2005.

\bibitem{NIST:2016}
\BIBentryALTinterwordspacing
E.~Barker, ``{Recommendation for Key Management Part 3: Application-Specific
  Key Management Guidance(Revision 4 - NIST Special Publication 800-57)},''
  2016, retrieved at November 11, 2016. [Online]. Available:
  \url{http://nvlpubs.nist.gov/nistpubs/SpecialPublications/NIST.SP.800-57pt1r4.pdf}
\BIBentrySTDinterwordspacing

\bibitem{PFUBLMPSKWHJZ:2004}
A.~Poppe, A.~Fedrizzi, R.~Ursin, H.~B{\"o}hm, T.~Lor{\"u}nser, O.~Maurhardt,
  M.~Peev, M.~Suda, C.~Kurtsiefer, H.~Weinfurter, T.~Jennewein, and
  A.~Zeilinger, ``{Practical quantum key distribution with polarization
  entangled photons},'' \emph{Optics Express}, vol.~12, no.~16, pp. 3865--3871,
  2004.

\bibitem{IANA:2012}
\BIBentryALTinterwordspacing
{Internet Assigned Numbers Authority}, ``{IPSEC ESP Transform Identifiers},''
  2012, retrieved at November 11, 2016. [Online]. Available:
  \url{http://www.iana.org/assignments/isakmp-registry/isakmp-registry.xhtml\#isakmp-registry-9}
\BIBentrySTDinterwordspacing

\bibitem{KJSHL:2013}
J.~Kang, K.~Jeong, J.~Sung, S.~Hong, and K.~Lee, ``{Collision Attacks on
  AES-192/256, Crypton-192/256, mCrypton-96/128, and Anubis},'' \emph{Journal
  of Applied Mathematics}, vol. 2013, p. 713673, 2013.

\bibitem{RFC4308}
P.~Hoffman, ``{Cryptographic Suites for IPsec},'' Internet Requests for
  Comments, {Internet Engineering Task Force}, {RFC} 4308, 2005.

\bibitem{kim2008birthday}
J.~H. Kim, R.~Montenegro, Y.~Peres, and P.~Tetali, ``A birthday paradox for
  markov chains, with an optimal bound for collision in the pollard rho
  algorithm for discrete logarithm,'' in \emph{International Algorithmic Number
  Theory Symposium}.\hskip 1em plus 0.5em minus 0.4em\relax Springer, 2008, pp.
  402--415.

\bibitem{McGrew:2012}
D.~A. McGrew, ``{Impossible plaintext cryptanalysis and probable-plaintext
  collision attacks of 64-bit block cipher modes.}'' \emph{IACR Cryptology
  ePrint Archive}, vol. 2012, p. 623, 2012.

\bibitem{ECRYPT:2012}
\BIBentryALTinterwordspacing
``{ECRYPT II Yearly Report on Algorithms and Keysizes (2011-2012)},'' 2012,
  retrieved at November 11, 2016. [Online]. Available:
  \url{http://www.ecrypt.eu.org/ecrypt2/documents/D.SPA.20.pdf}
\BIBentrySTDinterwordspacing

\bibitem{FRBS:2004}
E.~Freeman, E.~Robson, B.~Bates, and K.~Sierra,
  \emph{\BIBforeignlanguage{English}{{Head First Design Patterns}}}.\hskip 1em
  plus 0.5em minus 0.4em\relax Sebastopol: O'Reilly, 2004.

\bibitem{vdLinden:1994}
P.~{von der Linden}, \emph{\BIBforeignlanguage{English}{{Expert C Programming:
  Deep C Secrets}}}.\hskip 1em plus 0.5em minus 0.4em\relax Upper Saddle River:
  Prentice Hall, 1994.

\bibitem{Maur:2015}
O.~Maurhart and C.~Pacher, ``{AIT QKD R10 Software},'' 2015,
  \url{https://sqt.ait.ac.at/software/projects/qkd}, (accessed: Feb.26, 2016).

\bibitem{Vernam:1926}
G.~S. {Vernam}, ``{Cipher Printing Telegraph Systems For Secret Wire and Radio
  Telegraphic Communications},'' \emph{{Transactions of the American Institute
  of Electrical Engineers}}, vol. XLV, pp. 295--301, 1926, reprint B-198.

\bibitem{varga2001omnet++}
A.~Varga \emph{et~al.}, ``{The OMNeT++ discrete event simulation system},'' in
  \emph{{Proceedings of the European simulation multiconference
  (ESM{\rq}2001)}}, vol.~9, no. S 185.\hskip 1em plus 0.5em minus 0.4em\relax
  sn, 2001, p.~65.

\end{thebibliography}

\end{document}